Tech Science Press

# Preserving Data Confidentiality in Association Rule Mining Using Data Share Allocator Algorithm

## D. Dhinakaran[1,*] and P. M. Joe Prathap[2]

[1]Department of Information and Communication Engineering, Anna University, Chennai, 600025, India
[2]Department of Information Technology, R.M.D Engineering College, Tiruvallur, 601206, India
*Corresponding Author: D. Dhinakaran. Email: dhinakaran@velammalitech.edu.in

**Abstract:** These days, investigations of information are becoming essential for various associations all over the globe. By and large, different associations need to perform information examinations on their joined data sets. Privacy and security have become a relentless concern wherein business experts do not desire to contribute their classified transaction data. Therefore, there is a requirement to build a proficient methodology that can process the broad mixture of data and convert those data into meaningful knowledge to the user without forfeiting the security and privacy of individuals' crude information. We devised two unique protocols for frequent mining itemsets in horizontally partitioned datasets while maintaining privacy. In such a scenario, data possessors outwork mining tasks on their multiparty data by preserving privacy. The proposed framework model encompasses two or more data possessors who encrypt their information and dispense their encrypted data to two or more clouds by a data share allocator algorithm. This methodology protects the data possessor's raw data from other data possessors and the other clouds. To guarantee data privacy, we plan a proficient enhanced homomorphic encryption conspire. Our approach ensures privacy during communication and accumulation of data and guarantees no information or data adversity and no incidental consequences for data utility. Therefore, the advantages of data mining have remained redesigned. To approve the exhibition of our protocols, we implemented the protocols through broad experiments, where the assessment outcome showed that the mined results obtained by our protocols are reliable to those obtained by a traditional sole-machine approach. Meanwhile, the findings of our performance assessment have shown that our methodology is very efficient, with reasonably reduced communication time and computation costs.

## 1 Introduction

Data mining (DM) is an innovation that investigates data concealed in mass information, and association rule mining is a one-off exemplary algorithm in DM. Association rule mining can discover the significant relationships between things from the information base. When the cloud server (CS) gathers different information, it invokes a mass of sensitized data of clients, and the exposure of such sensitized data might hurt clients' security. Subsequently, it is imperative to secure private data during the DM interaction. Since information security stands out enough to be noticed, different scientists have made many examinations on PPDM [1]. Several privacy-preserving ARM proposals have recently been considered based on homomorphic encryption [2]. Cryptography has a significant impact on our everyday lives, albeit numerous public is not entirely aware of the reality. Cryptography ensures our touchy data. On the off chance that some arbitrator catches enciphered text, it ought to be essentially unrealistic to think about the data processor's information. If we update the keys from time to time, we want to provide the highest level of security. The question emerges, how to switch over the key before safely switching over the authentic information. In our futuristic times, the demand for privacy is ever-increasing, and homomorphic encryption is one of the essential strategies to help save. Homomorphic encryption permits us to do more procedures on enciphered information with no earlier information about the private key and without unraveling that information, accomplishing the motivation behind securing information protection and finishing DM. It is critical to parallelize data mining methods used on distributed data. Parallelization solutions for DM calculations are vital to executing distributed and collaborative memory systems. The Parallel framework is a combination of low-cost parallel computations. The clearest contention for parallelism spins around information base size. Parallel processing is to improve the effectiveness of the DM algorithm. What makes parallelism suffering is the consistent upgrade in the calculation speed of processors. Accordingly, using security processors and various calculations of achievement of Association Rules is consistently an advantage. Information Parallel circulates the information over the accessible processors.

Various models are used to increase information security and protection. In this article, we spotlight privacy-preserving on a dispersed dataset. The framework replica has included two data processors and at least two clouds. Each data processor has its data collection, which they encrypt before outsourcing to the cloud worker [3–5]. The cloud is in charge of acquiring and storing information bases from various data processors, identifying frequent itemsets for data processors, and communicating mining findings to the appropriate data processors. For steady enhancements in the computing rate of the cloud, the diminished expense of the process, and the load-bearing, parallel processing approach will be the best solution. Using similar clouds and data share allocator algorithms to achieve Association Rules is continually an advantage. We segment the dataset into records in the information parallel process and distribute non-overlapping sets to each available cloud.

We construct a model with two protocols named protocol A and protocol B. In protocol A, each cloud accepts a dataset block from the data processor, searches for frequent itemsets, encrypts the block result using the enhanced homomorphic encryption algorithm, and sends the encoded block result to the master cloud combines all of the cloud block results. In protocol B, clouds dispatch their block outcome to all data processors to aggregate each cloud block outcome. The fundamental goal of our work is to ensure the privacy of the information base, miners, and mining results, to permit data processors and miners to go offline, to decrease the measure of calculation and traffic. In the second part, we discuss the related work. In the third part, we present our parallel computation of the ARM approach without compromising privacy. The fourth part discusses our two protocols, analysis by master cloud and process by data possessor. The fifth portion presents the security study and comparisons of our propounded two protocols for the association rules mining approach. Finally, we conclude with a conclusion.

## 2 Related Work

Data mining innovation is a cross-disciplinary study that comprises a set of data analysis tools, such as statistics and artificial intelligence, and a wide range of privacy-preserving (PP) strategies. Cryptography techniques, anonymization, obfuscation, and Secure Multiparty Computation are among the keys for PPDM. However, these solutions do not have many impediments. Accurateness of the data scrutiny on changed data is diminishing, high communication time and calculation costs, Original data values cannot be recreated, low productivity, and fading the satisfactoriness of the principle data.

Zhang Fengli et al. [6] focused on medical privacy, and they did not utilize the conventional reproduction algorithm to remake another dataset. Instead, they further developed the traditional recreation-based PPDM, proposed another underlying model for medical data transmission with privacy-preserving, and proposed a sanitization function for the sensitized rules masking. Rongxing Lu et al. [7] aimed to provide outsourced database solutions for multiple data owners. So that they can confidently share their data without sacrificing data confidentiality for vertically divided databases, they proposed a well-organized symmetric homomorphic encryption technique. Furthermore, they proposed a secure comparison approach to ensure that supports/confidences are made safe. In addition, they offered a cloud-assisted frequent itemset exploring a solution to form an association rule.

M Dhage et al. [8] tended by incorporating the horizontal partitioning concept to split the database into two separate halves and the rule generation process, improving data privacy. Using the FP Growth measure, they expanded their work for relational databases. Lin LIU et al. [9] resolved the issue of data privacy for Several Parties on Outsourced Cloud Data. They proposed a PP-ARM system where data communication occurs from several sites in a twin-cloud-designed structure. They developed a group of cryptographic blocks for PP-ARM drew on the BCP cryptosystem. Their approach draws on a group of sophisticated bilateral reliable computation algorithms. Jıe Chen et al. [10] endeavor the privacy anxiety by utilizing the LWLR process. They process a Privacy-preserving privately distorted linear regression scheme system, where their approach encrypts the best-fit curve and protects users' privacy. They utilized Paillier homomorphic cryptogram to encipher data and afterward applied the arbitrary descent in the enciphered domain.

F Wang et al. [11] addressed the security cases and how to progress the precision of online medical analysis. With a privacy-preserving cooperative model acquisition scheme, healthcare organizations may firmly discover global process models with their local evaluation methods in the cloud, and each healthcare organization's sensitive medical data is effectively protected. Nankun Mu et al. [12], to mitigate the deficiency of data effectiveness they proposed a data set remaking-based calculation for Privacy-preserving frequent itemset mining (PPFIM) that can accomplish a severe level of privacy as well as bear the cost of a sensible data utility. A specialized sanitizing technique in database recreating - PPFIM - all delicate frequent itemsets with linked frequent itemsets detected for deletion in the pre-clean cycle. They proposed a creative database reproduction technique to reconstruct a correct information base using the frequent itemsets.

Jun Liu et al. [13] pointed out the two most normal conveyed DM errands that are not adhered to in multiparty computation (MPC) that are normal in reality. Two examples are calculating statistics for various data types and investigating the relationship between elements using linear regression. To help with the requirements in the MPC setting, they plan a calculation based on one-hot concealment and LU decay with optimized matrix computing. Furthermore, the author employed a calculation based on the SPDZ convention, a cross-stage multiparty calculation programming structure based on the Multi-Protocol SPDZ convention stack, as the base system to execute their solution. Jae-Seong Lee et al. [14] proposed a novel approach dependent on distance-based record connection and miniature total to find some concord among utility of data and privacy revelation through the DM of sensitive data with no identifiers. To enable open government data (OGD) data mining by combining two de-recognized OGD into one dataset. Their strategy permits clients to change the protection edge level to decide between privacy divulgence hazard and information utility. To assess the adequacy of the technique by employing broad experimentation.

In this paper, we concoct a compelling methodology for privacy-preserving ARM. Our methodology finds an association rule in a dispersed environment with sensibly decreased communication time and computation costs, productively utilizing the data, high performance, preserves participants' privacy and gives precise results.

## 3 System Model

Our framework model emphasizes how to enable privacy-preserving collaborative model learning. To guarantee data privacy, each data possessor will do some pre-activities like information cleaning, data reduction, scrambling their data before sending it to an untrusted cloud. Furthermore, each data possessor can associate with other data possessors and the cloud. Unequivocally, the framework comprises of three sections: 1) data possessor (DP), 2) Cloud server - Block Mining (CBM), and 3) Cloud server - Consolidate Block Result (CCBR), as present in Fig. 1.

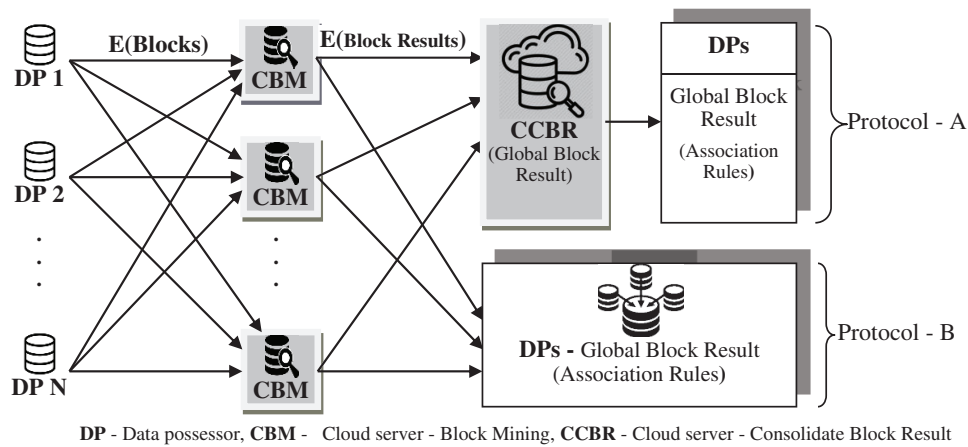

DP - Data possessor, **CBM** -   Cloud server - Block Mining, **CCBR** - Cloud server - Consolidate Block Result

**Figure 1:** System model – privacy preserving of DP's data through parallel computing

To collaborate with each data possessor to get a coordinated mining result, as an initial stage of the process, each data possessor encrypts their data by classical encryption technique to guarantee the privacy of their data. The traditional encryption method lessens computation time and cost than late encryption techniques [15–19]. Therefore, we can increase the number of rounds for traditional encryption to ensure protection. We embraced vigenere cipher for encoding data sets of each data possessor. Once data sources are encrypted, to partition the encrypted data sources, we use a data share allocator algorithm, which partitions the encrypted data source into multiple blocks and appropriates among the CBM. Block shamble & mapper are responsible for block mapping. Hence at each time of the entire process, the blocks are assigned to different CBM to ensure privacy. Each CBM is a cloud server that does the ARM for their resultant block and recognizes the frequent itemsets from the enciphered block allocated by the data share allocator algorithm. We proposed two protocols named protocol A and protocol B.

In protocol A, each CBM encrypts the block result using enhanced homomorphic encryption (EHE) and sends their enciphered block result to CCBR to cumulative all the CBM block results. In protocol B, CBM sends their block result to all data possessors to cumulative all the CBM block results. CBM initiates the maximum computations during our model's community-oriented model learning task. CCBR is a cloud server conscientious of amassing the enciphered block results and producing the concluding global diagnosis model *via* cooperative computing with CBMs. Finally, the overall output is made available to all data owners. The data possessor deciphers the global result to view the frequent itemset by utilizing the secretive key sent by CBM.

### 3.1 Design Goals

Our planned framework means to accomplish the accompanying five objectives:

1) The framework permits data possessors to discard complicated computing processes.

2) The framework permits parallel block computation, which diminishes the overall computation time.

3) The framework gives the outrageous level of privacy to all data possessors without uncovering any data possessor's private information, intermediate outcomes, and Mined results (Association Rule).

4) Because a trustworthy CS generates public and secret keys, the framework considers the secretive key disclose crisis.

5) The model allows for more flexibility. For example, the framework will not collapse when many parties are involved.

### 3.2 Preprocessing Stage

#### 3.2.1 Encrypting DP Data Sources:

The data resources of each DP are secured using a standard encryption approach to protect their privacy [7]. This paper adopts a vigenere cipher for encrypting data sets of each DP's. The finest and one of the uncomplicated, polyalphabetic ciphers is the vigenere cipher. In this method, the succession of Item text letters $I = I_0, I_1, I_2, \ldots, I_{n-1}$ and a key incorporate the succession of letters $SK = sk_0, sk_1, sk_2, \ldots, sk_{t-1}$, where typically transaction, $t < n$. The succession of encrypted letters $En = En_0, En_1, En_2, \ldots, En_{n-1}$ is deliberate as follows:

$$En = En_0, En_1, En_2, \ldots, En_{n-1} = E (SK, I)$$

$$= E [(sk_0, sk_1, sk_2, \ldots, sk_{t-1}), (I_0, I_1, I_2, \ldots, I_{n-1})]$$

$$= (I_0 + sk_0) \bmod 26, (I_1 + sk_1) \bmod 26, \ldots, (I_{t-1} + sk_{m-1}) \bmod 26, (I_t + sk_0) \bmod 26,$$

$$(I_{t+1} + sk_1) \bmod 26, \ldots, (I2_{t-1} + sk_{t-1}) \bmod 26, \ldots \qquad (1)$$

The key's principal characters are identical to those of the Item text characters, mod 26, adding subsequent Item characters, and so on, through the Item text's principal m characters. Next, we rehash the "key" letters for the successive t letters of the Item text character. This cycle proceeds until enciphering the entirety of the exchanges of a data set [20]. Finally, encipher every Item text character with an alternating cipher character in light of the "key" character's result.

An overall form of the encoding and decoding measure is

$$En_i = (I_i + k_{i \bmod t}) \bmod 26 \qquad (2)$$

$$I_i = (En_i - k_{i \bmod t}) \bmod 26 \qquad (3)$$

A key as long as the item text is necessary to encrypt data transactions. For the most part, the key is a duplicated keyword. For instance, if the keyword is medical, the item text "throat pain" is enciphered as "fluwctammq" as displayed in Tab. 1. To guarantee more protection of information sources, we can improve the encryption process for N rounds. Therefore, the key will be changed to secure the information source.

**Table 1:** Pre-processed datasets

| PID | Disease | PID | Disease |
| --- | --- | --- | --- |
| P1 | cold, cough, body pain | P1 | osol, osxoj, nsggratz |
| P2 | cold, fever, cough | P2 | osol, riymt, osxoj |
| P3 | throat pain, cold, fever | P3 | fluwctammq, osol, riymt |
| P4 | cold, cough | P4 | osol, osxoj |
| P5 | cold, fever, cough, body pain | P5 | osol, riymt, osxoj, nsggratz |
| P6 | cold, fever, cough, body pain, throat pain | P6 | osol, riymt, osxoj, nsggratz, fluwctammq |
| … | ……, ………, ……….. | … | ……, ………, ……….. |

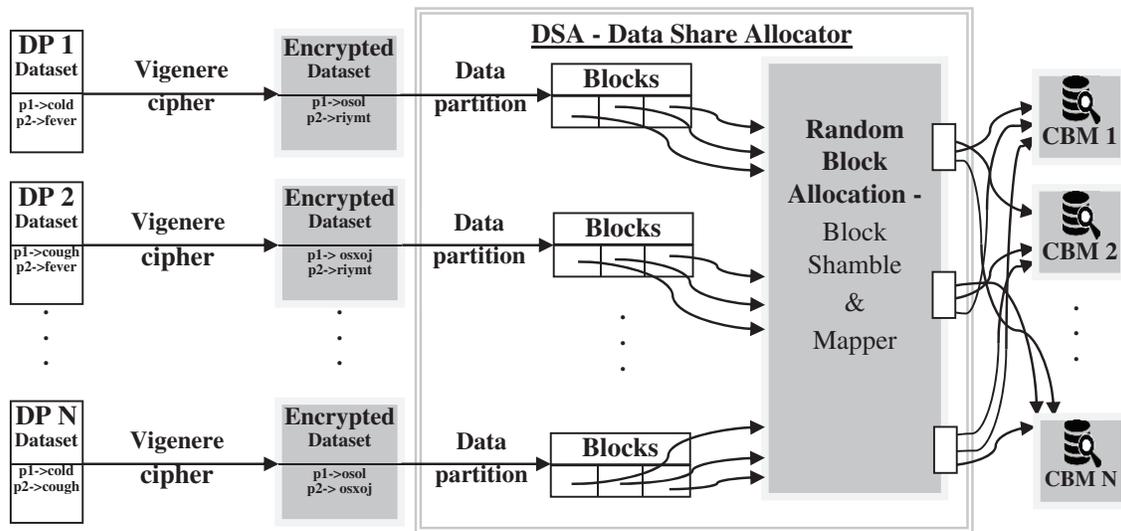

**Figure 2:** Pre-processing model using DSA algorithm

### 3.2.2 Data Share Allocator Algorithm:

A data share allocator algorithm is a horizontal data partition that encloses a subset of the whole information source and henceforth is conscientious for serving a CBM of the general responsibility. The data share allocator algorithm has two significant functionalities, data partitioner and block shamble & mapper as shown in Fig. 2. Data partitioner functionality is to level out an information source through horizontal fragmentation. A horizontal fragment of an information source is a subset of the tuples in that connection. A condition on at least one attribute of the relationship determines which tuples have a position with the horizontal fragment. Data partitioner helps in versatility, accelerating handling, and geo-conveyance by horizontally partitioning information. The following functionality is block shamble & mapper, which produces an arbitrary block allocation based on N CBM $(1 \leq i < N)$. It creates an arbitrary number in O (1) time. Working of algorithm flinches from the last block; swap it with an arbitrarily chosen block from the entire list of the block.

### Data Partitioner:

As presented in Fig. 3, data partitioner executions utilize dependable hashing for appropriating an information base consistently across the CBM in the topology. A partitioner key recognizes every information item in the data resource. Partitioner keys have been produced arbitrarily by data possessors for each tuple in the data resource to keep from design location. We were hashing the partitioner keys into a block. Each block addresses a part of the data resource, alluded to as virtual shares. Data partitioner traces between the apportioning "key" values and its CBM depending on the consequence of a hashing task. Partitioner takes a share key-value and produces a hash value from it. Using the hash number to determine which block the information transaction should reside. With a uniform hashing calculation, the hash function can equitably circulate information across blocks, reducing the risk of hotspots [21]. Consummate by processing a numeric hash number out of the key and calculating the modulo of that hash utilizing the total amount of CBM to figure which block claims the key. With this methodology, information transaction with close share keys is almost certainly not obtainable to put on a similar CBM. This design is hence excellent for designated information activities.

Share Block value $=$ Partitioning key values % Total number of CBM                              (4)

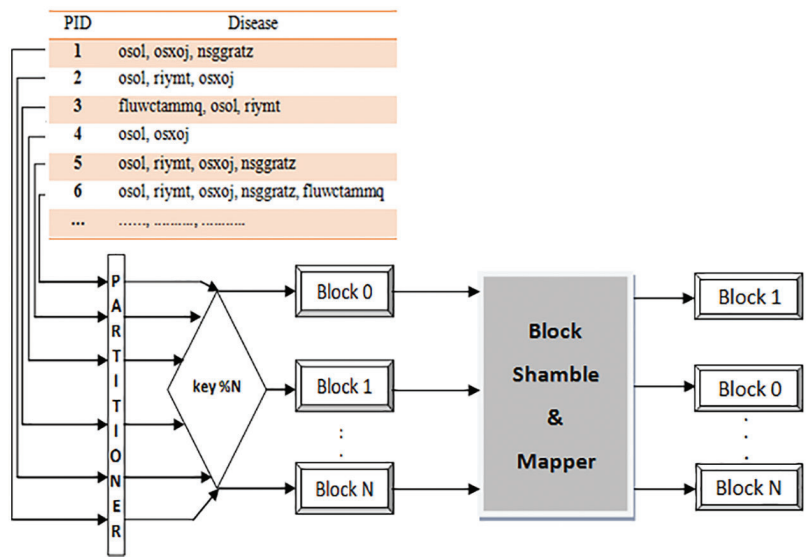

**Figure 3:** Process of data share allocator

*Block Shamble & Mapper:*

Block shamble & mapper considers the list of blocks from 1 to N-1 (size diminished by 1) and replicates the process till we knock out the first block. A significant property of the block shamble algorithm is that each block has an equivalent likelihood, of 1/N, of being picked as the last block in the block list. End of shambling, we will get a well-shuffled list of blocks. As a result, blocks will map in succession to a list of CBM. For instance, say B1, B2, B3, and B4 are the four blocks associated with the process. Consider the output of the block shamble algorithm as B4, B1, B2 & B3. The blocks allocation will be B4 to CBM1, B1 to CBM2, B2 to CBM3, and B3 to CBM4.

---

**Data Share Allocator algorithm:**

**Data Partitioner:**

**Input:** Key K, Data transaction T, List of CBM N

**Output:** A-List, L of N blocks indexed from 1 to N-1

  blockIndex = blockCode(K)

  blockIndex % = N

  Shareblock[blockIndex] = T

**Block Shamble & Mapper(L):**

**Input:** A-List, L of N blocks indexed from 1 to N-1

**Output:** A permutation of L so that all permutations are equally likely

  Ford = N-1 downwards to 1 do

    Let j←random (d+1)   // j is a arbitrary integer in [1, d]

    Swap L[d] and L[j]  // swap L[d] with itself, if j = d

return L

---

Emphasizing, for every one of the blocks that have not been shamble yet, initiate from the last array and move reverse throughout each block. At the last of the array, all swaps happen on-site, and when a block has been "shambled" into place, it cannot move once more. It restricts the feasible outcomes to equivalent to the digit of mix that is conceivable.

## 4  Mining Process

### 4.1  Protocol - A: Computation Based on CCBR

As presented in Fig. 4, each CBM does the three significant activities: calculating the block allocated by the data possessors, generating keys, encrypting the computed block result, and sharing the key with all data possessors. For computing the block, CBM makes use of the apriori algorithm to compute the frequent itemset and the association rule for the assigned block. Since the block is encrypted using classical encryption, CBM will not know the data possessor's data source. By taking advantage of classical encryption, the mining process does not affect, since no keys are involved, and the encrypted data are still in the form of alphabetic characters. Applying normal computation of apriori algorithm yields to the encrypted frequent itemset and the Association rule for the particular assigned block by each DP. A ⇒ B, where A and B are two disjoint itemsets, expresses an association rule. A ⇒ B denotes that A's occurrence with a particular degree of certainty implies the occurrence of B in a similar transaction.

For example, we will use a medical database, where a transaction is a patient's infection list. If and only if Supp (A ∪ B) ≥ Ts and Conf (A ⇒ B) ≥ Tc, A ⇒ B is considered an association rule. We characterize Conf (A ⇒ B) as the confidence of A ⇒ B. The latter is the possibility of B's occasion given A's occasion (*i.e.*, Conf (A ⇒ B) = Supp (A ∪ B)/Supp (A)). Tc means the threshold confidence, and Ts indicates the threshold support.

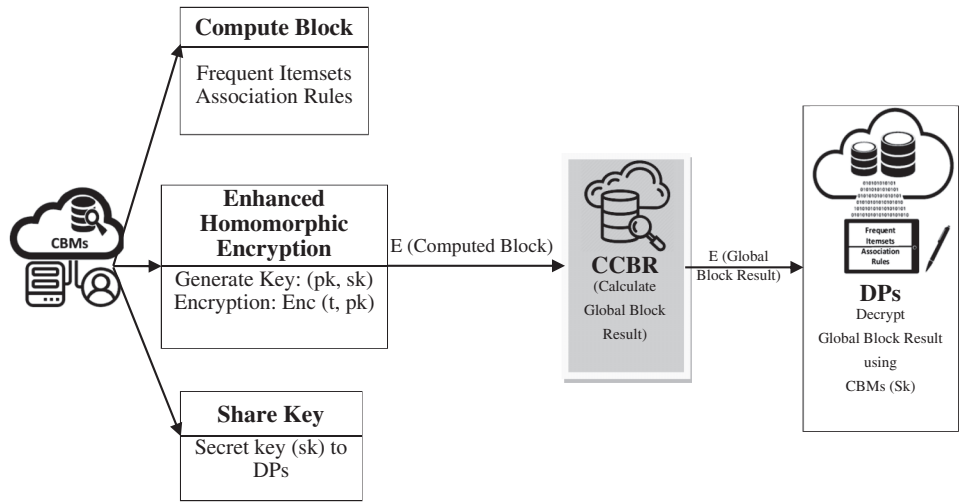

**Figure 4:** Computation based on CCBR

### 4.1.1 Enhanced Homomorphic Encryption on Computed Block Result:

Homomorphic encryption frameworks accomplish encrypted data functions without knowing a deliberate secret key; the DP is the solitary owner of the secret key. When we decrypt the operation result, it is equivalent to having approved the calculation on the raw data. For example, if E(x+y) obtain by adding E(x) and E(y) with no conscious knowledge of the values of x and y, additive homomorphism occurs. Homomorphic encryption comprises the following three phases:

*CBM Generates Keys - KeyGen( ):*

$$\text{KeyGen}(\lambda) \rightarrow (a, \ p, \ q) \tag{5}$$

Key generation is a probabilistic calculation, which accepts input as $\lambda$ a protective parameter and outputs public key $P_k = p$ and secret key $S_k = (a, q)$. $p \gg q$, where both p and q are to pick as large primes. The bit range of q depends on the protective parameter, and a will be an arbitrary whole number from $Z^*_p$.

*CBM Encrypts the Computed Block – E ( )*

Encrypt function is the algorithm that accepts secret key Sk, a block result b, where $0 \le b < n$ and a parameter e as inputs and outputs encrypted block result Eb ← E (Sk, b, e). The parameter e is a positive integer where $0 \le e < n$ is called ciphertext degree. Let ra signify a largest arbitrary positive integer, and the bit range of ra, |ra|, fulfills |ra| + |p| > |q|. Eb's irregular constituent is ra. The encryption of block result b is E (b).

$$E \ (Sk, \ b, \ e) \ = \ a^e \ (ra.q \ + \ b) \ \text{mod} \ p \tag{6}$$

*DP Decrypts the Global Block Result – D ( )*

Decrypt task is a deterministic calculation that accepts encrypted block result Eb and as inputs, the CBM provides a secret key Sk and the degree of the ciphertext e. Decrypt task outputs a block result b ← D (Sk, Eb, e). Let $s^{-d}$ indicate the multiplicative converse of $s^d$ in the field $F^p$. The accuracy verification of the decryption calculation is as shown below.

$$D \ (Sk, \ Eb, \ e) \ = \ (Eb \ \times \ a^{-e} \ mod \ p) \ mod \ q \tag{7}$$

$$= \ ((a^e \ (ra.q \ + \ b) \ mod \ p) \ \times \ a^{-e} \ mod \ p) \ mod \ q$$

$$= \ (ra.q \ + \ b) \ mod \ q$$

$$= \ b$$

### 4.1.2 Computing Global Block Result- CCBR

Evaluate function is an algorithm that takes encrypted blocks $Eb1, Eb2, \ldots, Ebn$ and outputs a global block result. For the encrypted block of $b1 + b2 \ mod \ q$, it is possible to calculate a modular addition of $Eb1$ and $Eb2$ if $e1 = e2$. To accurately decrypt $b1 + b2$ from its encrypted block, $(ra1 + ra2) \ q + b1 + b2 < p$ should be fulfilled. We pick the bit ranges of p, q, and arbitrary elements to guarantee that every enciphered block in our PP mining results deciphered accurately.

$$Eb_1 \ + \ Eb_2 \ mod \ p$$

$$= \ a^{e1} \ (ra1.q \ + \ b1) \ mod \ p \ + \ a^{e2} \ (ra2.q \ + \ b2) \ mod \ p$$

$$= \ a^{e1} \ ((ra1 \ + \ ra2) \ q \ + \ b1 \ + \ b2) \ mod \ p \ if \ e1 \ = \ e2 \tag{8}$$

## 4.2 Protocol- B: Computation Based on Data Possessor:

Each CBM does the two significant actions here, computing the block allocated by the data possessors and sending the computed block result to the entire data possessor. For computing the block, CBM makes use of the apriori algorithm to compute the frequent itemset and the association rule for the assigned block. Since the data possessor encrypts the block, CBM will not know the data possessor's data source. However, due to classical encryption, the mining process does not affect. Each data possessor obtains the encrypted frequent itemset and the association rule for the assigned block using the apriori algorithm's standard calculation [22–24].

### 4.2.1 Computing Global Block Result- Data Possessors

Each CBM has computed once frequent itemset and the Association rule for the assigned block, the outcomes of all CBM sent to the entire data possessor. Data possessors will aggregate [23] the frequent itemset and find the global Association rule from the computed block result received from all the CBM. We can have a data possessor portal and select a master data possessor who consolidates the frequent itemset and finds the global Association rule for better performance. At each procedure, a master data possessor chose at random. Once the master data possessor has identified the result, it will display it to all other data possessors through the portal.

$$Frequent \ Itemsets \ (FI) \ = \ \sum_{f=1}^{N} \ \sum_{CBM=1}^{n} \ \left( (FI)_f \right)_{CBM} \tag{9}$$

## 5 Security Analysis

This section looks at the security attribute of the recommended outcomes, specifically how our outcome might safeguard a data possessor's data out from arbitrator, CBM, and various data possessors.

## 5.1 Security Under the Third-Party / CBM Attacks:

CBM performs most of the calculations during collaborative model learning tasks, which is crucial in determining the Association rule. We used an utterly untrusted approach, in which CBM does the calculations in encrypted data. Data possessors will encipher the raw data to guarantee privacy. In our model, using vigenere cipher, we encrypt raw data. The vigenere cipher is subject to frequency analysis assaults. Therefore keep that in mind. Partitioning encrypts data, and each share/block is assigned to a separate CBM using an arbitrary allocation process to counteract such an assault. As a result, breaking one of these information blocks using item frequency analysis is unrelated to breaking another. CBM will have a piece of data possessor's enciphered information in light of the above pre-processing models. Henceforth, CBM has no idea about the information about data possessors [24–26]. Two protocols combine all CBM block results and determine the global block result. From Theorem 4 of [15], we know break likelihood for an itemset in such a data block is close to 1/k. Assume A is an itemset of the connected information block, with A's item(s) coming from d DP's. Then A could be separated with d itemset(s), all containing just one item from DP(s).

The attacker must shatter all d itemsets in order to break A. (s). The likelihood of breaking A is close to (1/k) d since breaking one of these itemsets frees the other itemsets. In our data, the break chance for an itemset is still close to 1/k as t ≥ d ≥ 1. In protocol A, CBM applies the EHE technique on the data share outcome and sends their encrypted block outcome to CCBR for the accumulation process. Due to the homomorphic scheme, CCBR will not have an idea of the block and the global result. In protocol B, CBM sends their block outcome to all data possessors for the accumulation process. Since the cloud server is not involved in this protocol, the global outcome will not reveal to CBM/CCBR. The PIDs may contain delicate data. To hide such information, the hash values of the primary PIDs use to restore the PIDs in the outward data block. Because of the cryptographic hash function's preimage resistance, the CBM cannot view the primary PIDs from the PIDs utilized in the outward data block.

### 5.2 Security Under Data Possessors' Attacks:

We adopted an outward DM process, where every data possessor performs a preprocessing process before sending them to CBM. Data possessors utilize protocol A or B to find the global frequent itemsets and association rule candidates. As a result, data possessors are not obliged to exchange plaintext or ciphertext with each other. Even data possessors have no permission to access others' ciphertext through CBM. If any data possessor colludes with any CBM to find the information of any other data possessor, our protocols prevent any information leaks. Furthermore, solid supports and confidences remain hidden from any data possessor due to the enhanced homomorphic encryption approach. Consequently, any data possessor cannot infer other data.

## 6 Performance Evaluation

We first examine the performance of the supplied protocols in terms of the computational complexity of our ARM and frequent itemset mining solutions in this section. Then, as a pattern, we apply one of [11]'s solutions and typical non-privacy-preserving strategies.

Our two protocols accomplish practically a similar privacy level. Conversely, other solutions accomplish lower security levels. Traditional non-privacy algorithms were used as baselines because they are currently more effective ways accessible.

### 6.1 Computation Cost Analysis:

#### 6.1.1 Comparing with Traditional Non-Privacy Algorithms:

To determine the computational complexity, we used the running time. We compare our methods to the most productive non-privacy-preserving algorithms currently available to demonstrate the practicality of our work. We evaluated our ideas and traditional algorithms (Apriori, FP-growth, and Eclat) utilizing medical scrutiny datasets got from "https://data.world/datasets/health." We segregated each dataset into t data sets arbitrarily to simulate t data possessor (DP). NetBeans is where we do our work. Our investigations utilize a JAVA execution of Apriori, FP- growth, and Eclat algorithms. In all investigations, six laptop computers playing the jobs of cloud and data possessors have similar hardware and software settings to guarantee a reasonable correlation. Uncommonly, we use two laptops as the data possessors, which encipher the information and transfer it to the cloud. The rest are utilized as the cloud and Evaluator separately.

Tab. 2 shows the experimental parameters for our solutions. The outcomes (running time) of ARM in Figs. 5–8. We divided the run time as cloud end (*i.e.*, mining time) and DP ends (for example, running time of preprocessing and decoding). Based on our findings using various parameters (t, k, and c) and data sources, we show that the run time of each of our solutions is just one order of magnitude longer than the most acceptable non-PPDM algorithms. Resource utilization is extremely low in the DP end since both information and processing work shifts to the CBMs. Our solutions are as effective as the most advanced low-privacy techniques. In most cases, the CS's running time is roughly one order more than the traditional algorithm's, whereas the DP's running time is less than the traditional algorithm's.

**Table 2:** Experimental settings

| CPU | Intel(R) Core i5-7200U @ 2.50 GHz |
|---|---|
| Software | Windows 10 64bit and NetBeans |
| Memory | 8 GB |
| Data | World/datasets/health |
| Data bit length | < 72 bits |
| λ (security parameter) | 80 |
| \|ra\| for $Eb_s$, $Eb_e$ and ERVs | = 40 bits = λ/2 |
| \|ra\| for $Eb_z$ | 64 bits |

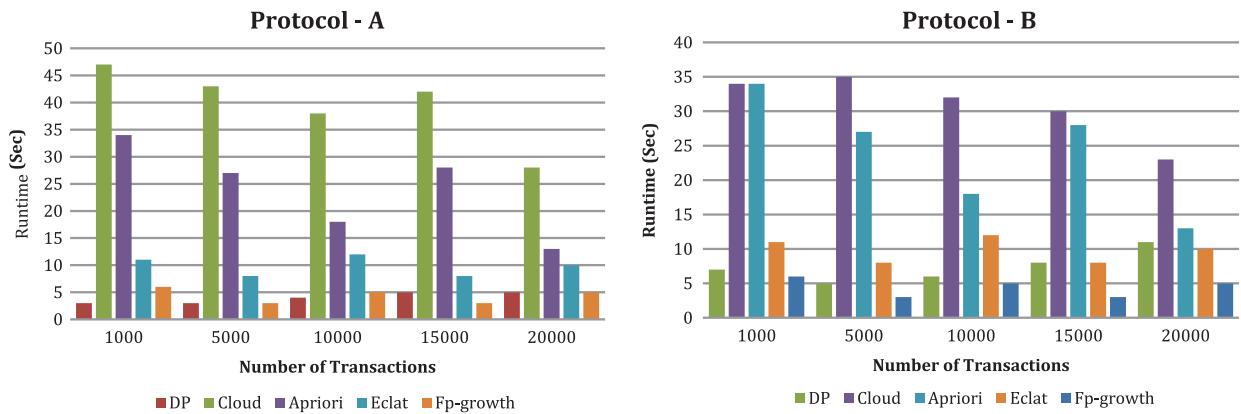

**Figure 5:** Runtime comparison (t = 3 and k = 8) with traditional non privacy algorithms

From Figs. 6 and 7, we likewise see that runtime changes with expanding upsides of k and t. With the criteria k and t, the CS's runtime increases. When k is set to 10 and scaling the criterion t from 3 to 9, we tracked down the runtime of CS increments. The runtime of DP diminishes in both conventions. When the number of DP is set to 3 and scaling the criterion k from 6 to 12, we tracked down that the runtime of CS increments and the runtime of DP changes marginally. From Fig. 8, we see that runtime decline with expanding the quantity of CBM, and the runtime of DP changes with expanding the number of CBM by setting the criterion t as three and the criterion k as 10. In our investigation, contrasted with protocol A, DP runtime is high in protocol B. We can likewise see that when t increases, DP's running time diminishes. The explanation is that if a similar joint database divides into more DP, every DP's dataset is more modest. Therefore, it takes less time to preprocess a smaller dataset. As a result, DP's running time does not increase as k increases. Expanding t and k, for the most part, yields a longer runtime at the cloud side but no increase in DP's running time when using our methods.

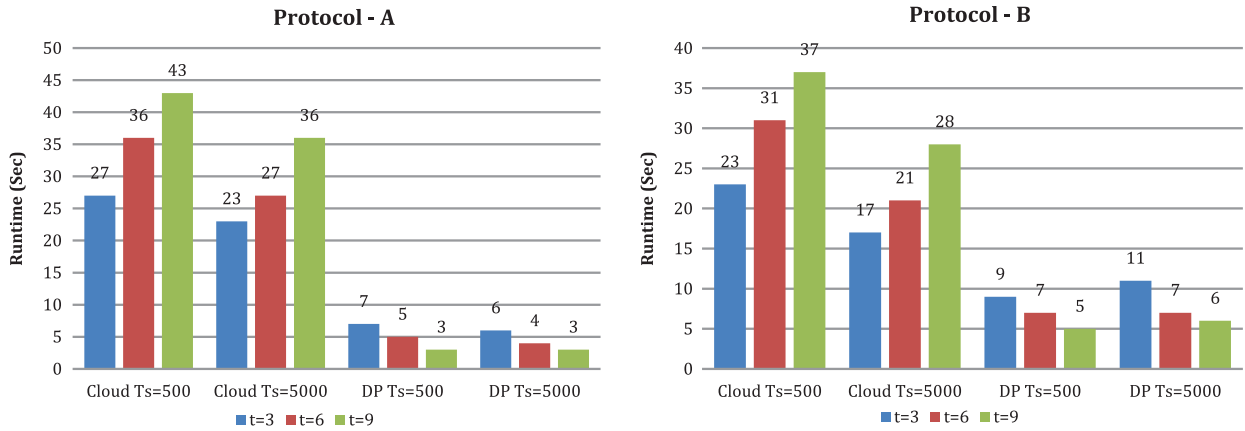

**Figure 6:** Runtime under diverse DP count t (k is set to 10)

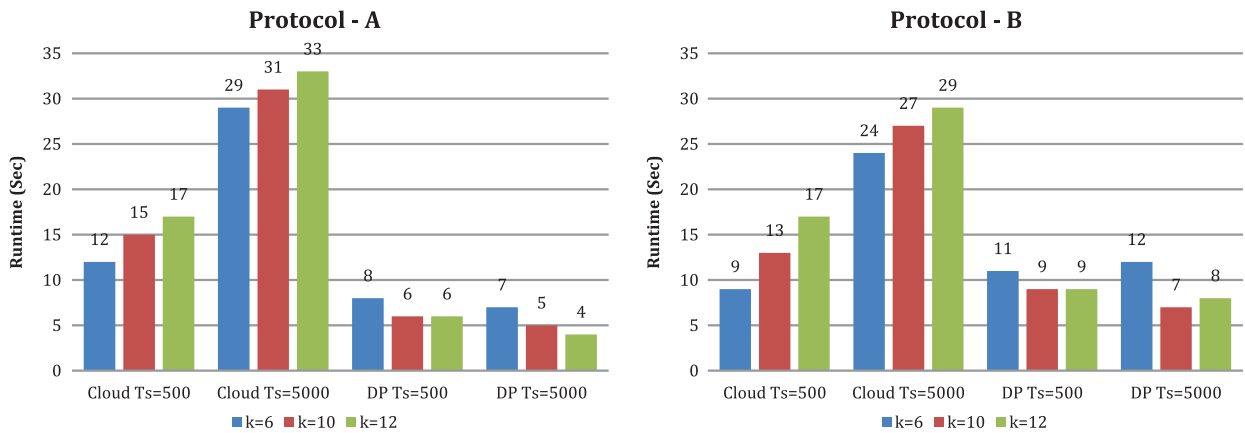

**Figure 7:** Runtime under diverse Itemset k (t is set to 3)

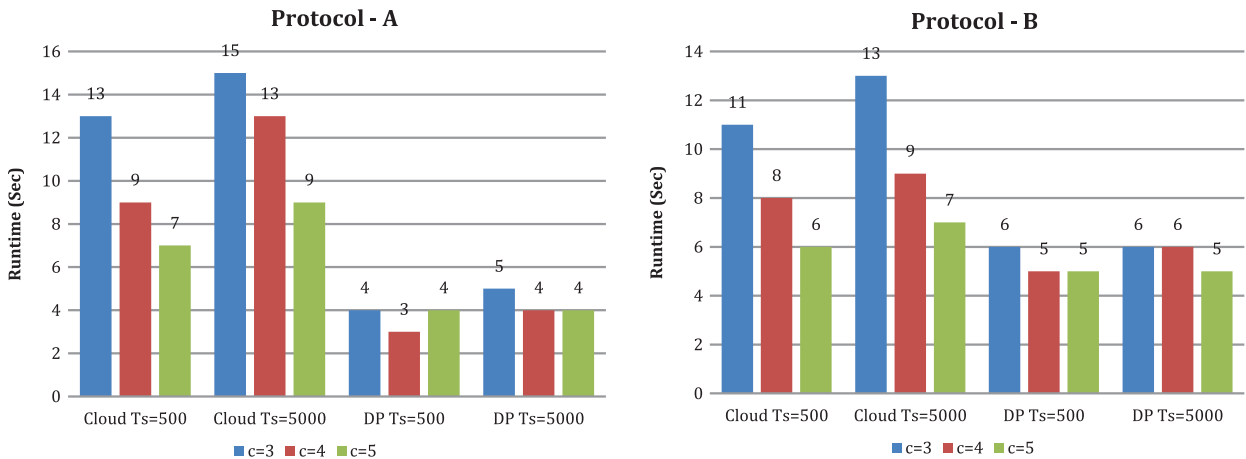

**Figure 8:** Runtime under multiple cloud c (t is fixed to 3 and k is fixed to 10)

*6.1.2 Comparing with Privacy Algorithms:*

Using the same arrangement of hardware, software, and datasets to execute contrasts with non-privacy Algorithms. The computation cost of our solution is a lot of substandard identified with the disinfected dataset, Cryptographic Hash Function, and outwork ARM. One of the most incredible existing privacy-preserving solutions that do not release sensitive data of the raw information is [11]'s frequent itemset mining utilizing privacy-preserving collaborative model learning(PCML) scheme.

Like our solutions, [11]'s solution utilizes a homomorphic encryption scheme for ARM and frequent itemset mining. Tab. 3 depicts the time requirements for executing the existing and proposed frameworks, as determined by the mathematical model. The execution time of this solution is more than the proposed framework because of its costly operations. Fig. 9 analyzes the computation cost of our methodology with a PCML scheme by fluctuating the number of transactions and the range of criteria (t,k,c) to track down the frequent itemsets. We set the criterion value as DP = 4, k = 8, and c = 3. We guarantee that our design and execution are versatile when the data size increments based on the outcome. we likewise see that the runtime of proposed protocols is practically comparative, and the runtime of proposed protocols is more diminutive in contrast with the PCML scheme.

**Table 3:** Runtime comparison

| System | Time in ms |
| --- | --- |
| Existing system (PCML) without horizontal segmenting with 20000 transactions | 72400 |
| Proposed system - Protocol A, with horizontal segmenting with 20000 transactions | 54700 |
| Proposed system - Protocol B, with horizontal segmenting with 20000 transactions | 47640 |

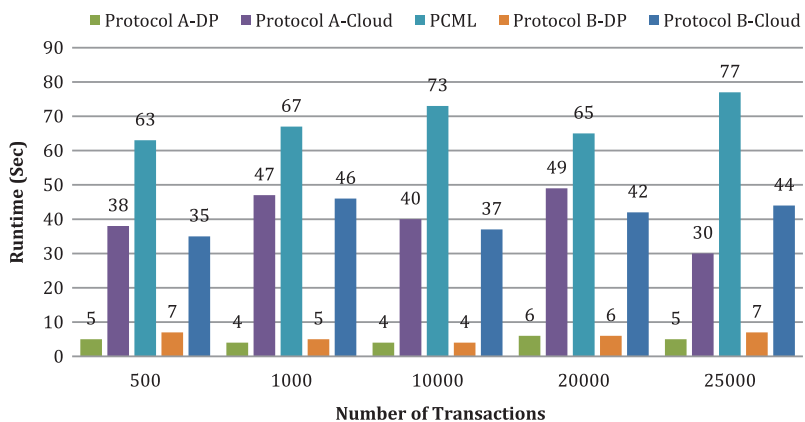

**Figure 9:** Computation cost analysis with privacy algorithm

# 7  Conclusion

Security protecting is an essential task for data mining. This article proposes two unique and practical techniques for frequent mining itemsets in horizontally partitioned datasets while maintaining privacy. This technique permits the data possessors to outsource data mining tasks on their shared data while maintaining anonymity. The proposed framework model incorporates at least two data possessors who encrypt their data using a classical encryption scheme. Furthermore, distribute their encrypted data to at least two clouds by utilizing the DSA algorithm. The framework uses two major functionalities, data partitioner, and block shamble & mapper, to partition the dataset based on records and map non-overlapping sets of records to each available cloud arbitrarily. Hence this approach protects raw data from other data possessors and the clouds. As an effect of parallelization, implementation speed of cloud improves continually; execution of the outwork ARM improves, the diminished expense of the course and balances the load.

We created privacy-preserving outsourced ARM results for horizontally partitioned databases. Our protocols protect the raw data of data owners from other data owners and the cloud. Our approach guarantees no information or data loss and with no incidental effects on data utility and benefits of DM has remained upgraded. Both protocols ensure privacy and data security in the communication and aggregation of data. We carried out the protocols during extensive experimentations to confirm the competence and measure our plan's performance and execution. The assessment outcomes have shown that the mining outcomes of our protocols are indistinguishable from those created by a standard single-machine approach. Furthermore, the results of the sufficiency review have confirmed that our strategy and execution are pretty effective, with reasonably reduced communication time and processing costs; therefore, for empirical data mining jobs, our protocols are extensively employed. Our protocols are especially well-suited to data possessors who wish to move their data source to the cloud while preserving utmost confidentiality without losing performance.

In order to recognize our solutions, this article presented two practical protocols: protocol A - Computation dependent on CCBR and protocol B - Computation dependent on Data Possessor. Beyond the data mining techniques outlined in this research, both prototypes have potential applications in more safe computing applications, such as secure data accumulation. Prospect research will be focused on representing the effectiveness of the suggested homomorphic encryption system in additional scenarios and further enhancing the efficiency of our scheme. In conclusion, we opt to expand our work from relatively basic numerical calculation to more artificial intelligence tasks.

**Funding Statement:** The authors received no specific funding for this study.

**Conflicts of Interest:** The authors declare that they have no conflicts of interest to report regarding the present study.